\input epsf
\documentstyle{mn}

\newif\ifAMStwofonts

\newcommand{\erg}{{\rm erg}\, {\rm cm}^{-2}\, {\rm s}^{-1}}
\newcommand{\SIS}{SIS0}

\ifoldfss
  \ifCUPmtlplainloaded \else
    \NewTextAlphabet{textbfit} {cmbxti10} {}
    \NewTextAlphabet{textbfss} {cmssbx10} {}
    \NewMathAlphabet{mathbfit} {cmbxti10} {} 
    \NewMathAlphabet{mathbfss} {cmssbx10} {} 
  \fi
  \ifAMStwofonts
    \ifCUPmtlplainloaded \else
      \NewSymbolFont{upmath} {eurm10}
      \NewSymbolFont{AMSa} {msam10}
      \NewMathSymbol{\upi}     {0}{upmath}{19}
      \NewMathSymbol{\umu}     {0}{upmath}{16}
      \NewMathSymbol{\upartial}{0}{upmath}{40}
      \NewMathSymbol{\leqslant}{3}{AMSa}{36}
      \NewMathSymbol{\geqslant}{3}{AMSa}{3E}

    \fi
  \fi
\fi 

\ifnfssone
  \newmathalphabet{\mathit}
  \addtoversion{normal}{\mathit}{cmr}{m}{it}
  \addtoversion{bold}{\mathit}{cmr}{bx}{it}
  \newmathalphabet{\mathbfit} 
  \addtoversion{normal}{\mathbfit}{cmr}{bx}{it}
  \addtoversion{bold}{\mathbfit}{cmr}{bx}{it}
  \newmathalphabet{\mathbfss} 
  \addtoversion{normal}{\mathbfss}{cmss}{bx}{n}
  \addtoversion{bold}{\mathbfss}{cmss}{bx}{n}
  \ifAMStwofonts
    \ifCUPmtlplainloaded \else
      \UseAMStwoboldmath
      \makeatletter
      \new@mathgroup\upmath@group
      \define@mathgroup\mv@normal\upmath@group{eur}{m}{n}
      \define@mathgroup\mv@bold\upmath@group{eur}{b}{n}
      \edef\UPM{\hexnumber\upmath@group}
      \new@mathgroup\amsa@group
      \define@mathgroup\mv@normal\amsa@group{msa}{m}{n}
      \define@mathgroup\mv@bold\amsa@group{msa}{m}{n}
      \edef\AMSa{\hexnumber\amsa@group}
      \makeatother
      \mathchardef\upi="0\UPM19
      \mathchardef\umu="0\UPM16
      \mathchardef\upartial="0\UPM40
      \mathchardef\leqslant="3\AMSa36
      \mathchardef\geqslant="3\AMSa3E
    \fi
  \fi
\fi 

\ifnfsstwo
  \DeclareMathAlphabet{\mathbfit}{OT1}{cmr}{bx}{it}
  \SetMathAlphabet\mathbfit{bold}{OT1}{cmr}{bx}{it}
  \DeclareMathAlphabet{\mathbfss}{OT1}{cmss}{bx}{n}
  \SetMathAlphabet\mathbfss{bold}{OT1}{cmss}{bx}{n}
  \ifAMStwofonts
    \ifCUPmtlplainloaded \else
      \DeclareSymbolFont{UPM}{U}{eur}{m}{n}
      \SetSymbolFont{UPM}{bold}{U}{eur}{b}{n}
      \DeclareSymbolFont{AMSa}{U}{msa}{m}{n}
      \DeclareMathSymbol{\upi}{0}{UPM}{"19}
      \DeclareMathSymbol{\umu}{0}{UPM}{"16}
      \DeclareMathSymbol{\upartial}{0}{UPM}{"40}
      \DeclareMathSymbol{\leqslant}{3}{AMSa}{"36}
      \DeclareMathSymbol{\geqslant}{3}{AMSa}{"3E}
    \fi
  \fi
\fi 

\ifCUPmtlplainloaded \else
  \ifAMStwofonts \else 
    \def\upi{\pi}
    \def\umu{\mu}
    \def\upartial{\partial}
  \fi
\fi

\title[Deep Hard X-ray source counts]
  {Deep hard X-ray source counts from a fluctuation analysis of  ASCA
  SIS images}  
\author[Gendreau, Barcons \& Fabian]
  {K.C. Gendreau,$^1$ 
  X. Barcons$^{2,3}$ and A.C. Fabian,$^3$\\
  $^1$NASA/Goddard Space Flight Center, Greenbelt, MD 20771, USA\\ 
  $^2$Instituto de F\'\i sica de Cantabria (Consejo Superior de
  Investigaciones Cient\'\i ficas - Universidad de Cantabria), 39005
  Santander, Spain\\
  $^3$Institute of Astronomy, Madingley Road, Cambridge CB3 0HA}

\date{8 October 1997}
\pagerange{\pageref{firstpage}--\pageref{lastpage}}
\pubyear{1997}

\begin{document}

\maketitle

\label{firstpage}

\maketitle

\begin{abstract}
An analysis of the spatial fluctuations in 15 deep ASCA \SIS\ images
has been conducted in order to probe the 2-10~keV X-ray source counts
down to a flux limit $\sim 2\times 10^{-14}\erg$.  Special care has
been taken in modelling the fluctuations in terms of the sensitivity
maps of every one of the 16 regions ($5.6 \times 5.6\, {\rm arcmin}^2$
each) in which the \SIS\ has been divided, by means of raytracing
simulations with improved optical constants in the X-ray telescope.
The very extended `side lobes' (extending up to a couple of degrees)
exhibited by these sensitivity maps make our analysis sensitive to
both faint on-axis sources and brighter off-axis ones, the former
being dominant.  The source counts in the range $(2-12)\times
10^{-14}\erg$ are found to be close to a euclidean form which
extrapolates well to previous results from higher fluxes and in
reasonable agreement with some recent ASCA surveys. However, our
results disagree with the deep survey counts by Georgantopoulos et
al. (1997). The possibility that the source counts flatten to a
subeuclidean form, as is observed at soft energies in ROSAT data,
is only weakly constrained to happen at a flux $<1.8\times
10^{-12}\erg$ (90 per cent confidence). Down to the sensitivity limit
of our analysis, the integrated contribution of the sources whose
imprint is seen in the fluctuations amounts to $\sim 35\pm 13$ per
cent of the extragalactic 2-10~keV X-ray background.

\end{abstract}


\section{Introduction}

In the soft X-ray band (0.5-2~keV), a combination of direct source
counts in shallow and deep surveys (Hasinger et al 1993,
Branduardi-Raymont et al 1994) and analyses of the spatial
fluctuations (Hasinger et al 1993, Barcons et al 1994) has determined
the source counts down to a level where more than 70 per cent of the
extragalactic X-ray background (XRB) is resolved into sources.  The
most remarkable feature of the soft X-ray source counts is the
existence of a break at a 0.5-2~keV flux of $2\times 10^{-14}\erg$
below which the approximate euclidean behaviour that holds at brighter
fluxes flattens considerably.  The reason for this flattening in the
source counts is the steep evolution of the broad-line Active Galactic
Nuclei (AGN) which stops at redshift $z\sim 2$.  The so-called
Narrow-Line X-ray Galaxies (NLXGs), which might in fact be powered by
obscured AGN, appear in increasingly large numbers at fluxes $\ll
10^{-14}\erg$ (McHardy et al 1997, Romero-Colmenero et al 1996,
Almaini et al 1996). In a recent study Hasinger et al (1997) and
Schmidt et al (1997) have cast some doubts on the reality of these
NLXGs, since in their complete identification of the sources found in
the Lockman Hole above a flux of $5\times 10^{-15}\erg$ no such
objects appear.  Moreover, Hasinger et al (1997) also discuss the
severe confusion problems for deep surveys carried out with the ROSAT
Position Sensitive Proportional Counter below that flux (where most of
the NLXGs are found by other surveys).  Although this
might certainly affect some of the source
identifications at very faint levels, the X-ray sources putatively
identified as NLXGs have harder spectra than the broad-line
AGN (Almaini et al 1996, Romero-Colmenero et al 1996).  This indicates
that regardless of their optical counterparts, these sources might
be relevant to higher energies.

At harder X-ray energies (2-10~keV), where a larger fraction of the
energy of the XRB resides (see, e.g., Fabian \& Barcons 1992), our
knowledge of the X-ray source counts is more limited. Until ASCA
became operational, all the data in that energy band was collected by
collimated field-of-view proportional counters with angular resolution
of degrees. The Piccinotti et al (1982) sample was the only really
complete sample of hard X-ray sources going down to 2-10~keV fluxes of
$3\times 10^{-11}\erg$. Below that flux, a GINGA high galactic
latitude survey (Kondo 1991) and the GINGA fluctuation analysis
(Butcher et al 1997, Hayashida 1989) extended the euclidean source
counts found in the Piccinotti et al sample down to a flux of $5\times
10^{-13}\erg$. The surface density reached by these studies amounted
to a few sources per square degree to be compared with the deepest
source counts in the soft band reaching about 1000 sources per square
degree.

Even with its limited angular resolution, ASCA (Tanaka, Inoue \& Holt
1994) has opened the possibility of making a very significant step
forward towards the determination of the 2-10~keV source counts at
faint fluxes.  Various surveys have been carried out which show that
the source counts do not deviate dramatically from an euclidean
extrapolation of the source counts at higher fluxes. Among those
surveys, the Large Sky Survey (LSS, Inoue et al 1996, Ueda 1995)
covers 6 $\deg^2$ with the GIS down to a flux limit of $1.5\times
10^{-13}\erg$ for direct source detection. The Deep Sky Survey (DSS,
Inoue et al 1996) and the ASCA follow-up of 3 deep ROSAT fields by
Georgantopoulos et al (1997) provide rather discrepant source counts
down to fluxes of $4\times 10^{-14}\erg$ and $5\times 10^{-14}\erg$
respectively. Indeed, the small solid angle sampled by the deepest
surveys results in significantly large statistical uncertainties in
the number of sources detected down to the completeness flux. Other
effects, like inhomogeneities in the distribution of sources or X-ray
variability of the sources (see Barcons, Fabian \& Carrera 1997 for a
discussion on the effects of this last issue on the source counts) can
also affect estimates of source counts when only a small area of
the sky is covered.

The analysis of the spatial fluctuations in the XRB has been often
used to both improve on the surveyed area and eventually to determine
the source counts down to fainter fluxes than can be achieved via
direct source counting.  Examples of the application of this method in
X-ray imaging data can be found in Hamilton \& Helfand (1987), Barcons
\& Fabian (1990), Hasinger et al (1993) and Barcons et al (1994) among
others.  These analyses have either predicted or confirmed the source
counts at faint fluxes with success.  The theoretical sensitivity
limit of this method corresponds to a flux level for which there is
about one source per beam, although in some cases photon counting
noise prevents that limit being reached.

In this paper we present a first fluctuation analysis of 15 high
galactic latitude deep pointings obtained with the SIS0 detector on
ASCA. Down to the sensitivity level of our analysis ($\sim 2\times
10^{-14}\erg$) we find no compelling evidence for a flattening in the
source counts. We also find our results (which cover a nominal area of
2 $\deg^2$) to be consistent with the LSS and the DSS within $1\sigma$ 
uncertainties. The Georgantopoulos et al (1997) survey,
however, appears to be above our estimate of the source counts at
$5\times 10^{-14}\erg$ by more than 3 sigma.

In section 2 we present the data, taken from the archive, which has
been used in the current analysis.  Section 3 is devoted to explain
how the distribution of spatial fluctuations is modelled with special
emphasis on the sensitivity maps for each pixel and other effects.
Section 4 presents the results of the fits to the distribution of
fluctuations and the implications for source counts.  In section 5 we
summarize our results and discuss briefly possible extensions of this work.

\section {The Data}

Our data sample has been built by using all the public ASCA images in the
archive which comply with a list of selection criteria.  These are:
galactic latitude in excess of $20 \deg$ (to avoid the effects of
galactic absorbing columns close to $10^{21}\, {\rm cm}^{-2}$ which
would affect the visibility of sources above 2 keV), useful exposure
time (once the data has been cleaned) in excess of 20 ks and no bright
or extended X-ray targets in the image. 

To avoid the additional degradation in the Point-Spread-Function (PSF)
that the Gas Scintillation Proportional Counter already adds to the
rather limited angular resolution of the X-ray telescopes it was
decided to carry out the first analysis on the CCD data only.
Furthermore, since one of these detector, the SIS0, appears to be more
stable and with a more predictable behaviour than the SIS1, only SIS0
data have been used.  We further restrict ourselves to data
taken in 4-CCD mode since other images taken in 2-CCD and 1-CCD mode
would add very little to our data sample.

\begin{figure}
\vbox to 0cm{\vfil}
\epsfverbosetrue
\epsfysize=210pt
\epsffile{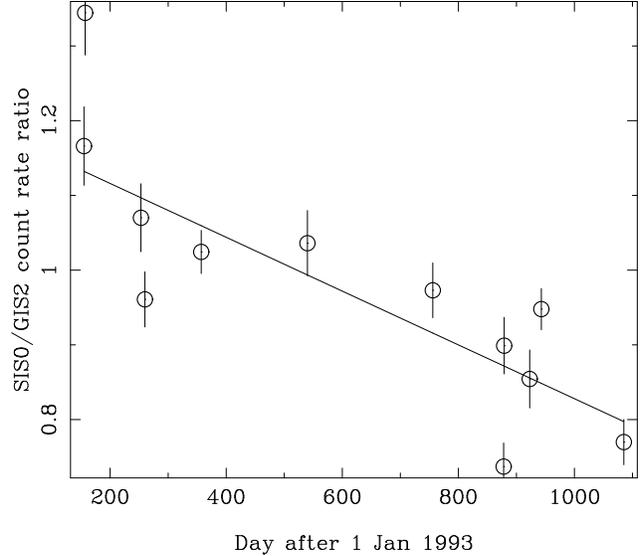}
\caption{Ratio of the SIS0 count rate to GIS2 count rate (in the same
area of sky seen by SIS0 and over the same energy band 2-7 keV) as a function
of date. The decreasing trend is the result of the RDD on
the CCDs.}
\end{figure}

There are also a few cases where two or more archival images partially
overlap.  In this case, we only use the deepest one, although through
a complicated process fluctuations in the `mosaiced' image could be
properly modelled. We believe that the additional effort in modelling
these very few images would make only a very small contribution to our
data sample. Table 1 presents the list of observations used.

\begin{table*}
  \vbox to 0cm{\vfil}
\centering
\begin{minipage}{22cm}
\caption{Details of the SIS data used}
\label{datalist}
\begin{tabular}{| l c c c c l |}
\hline Image name & RA (J200) & DEC (J2000) & $b_{II}$ &Days after & Exposure time \\
                  & (deg)     & (deg)       &  (deg) & 1-Jan-1993     &  (s)\\
\hline 
LYNX              & 132.30    &  -27.622     &  39 & 133      &  67260 \\
LOCKMAN1          & 163.00     & 47.176       & 53 &  144      &  25878 \\
DRACO1            & 256.30    &   -36.083    &   34 & 155      &  23128 \\
JUPITER           & 184.92    &  -0.594    & 61 &  157      &  24349 \\
QSF3 N2 N3 &  55.43   &  35.629   & -52 &  253      &  23301 \\
ANON              & 286.14    &   15.767   &  -15 & 260      &  24111 \\
IRAS10214         & 156.26     &  57.359      &  55 &  309      &  22864 \\
SA57 1            & 197.19    & 35.797      & 86 &  357      &  59038 \\
GSGP4             & 14.37    &   44.836    & -89 &  540      &  20128 \\
NGC1386           & 54.17    &  70.901    & -54 &  756      &  24076 \\
SA68              & 4.28    &   -26.697    & -46 &  923      &  28174 \\
QSO cluster       & 205.15    & 35.567      &  79 &  943      &  47367 \\
BRACCESSI1        & 194.32    &   -44.118    &   81 & 878      &  28206 \\
BRACCESSI2        & 196.18    &   27.317    &   81 & 879      &  24038 \\
BRACCESSI3        & 195.54    &   29.376    &  81 & 1085      &  26742 \\
\hline
\end{tabular}
\end{minipage}
\end{table*}

The first problem encountered in the analysis of data taken at such
different dates is that a form of the detector radiation damage (known
as the Residual Dark Difference -- RDD, see Dotani, Yamashita \&
Rasmussen 1995) does not affect all the data equally.  Indeed, the
CCDs have lost sensitivity with time and this effect has to be
properly accounted for in any fluctuation analysis.  Figure 1 shows
the count rate of the SIS0 detector for these data normalized to the
GIS2 count rate (over the same sky area covered by the SIS0 and for
the same energy band 2-7~keV) which is believed to be stable.  A very
clear trend of sensitivity loss is seen. For the purposes of the
fluctuation analysis, the standard model of a linear change in the
detector efficiency with time has been assumed. The scatter around the
model is moderate and is probably dominated by the brightest sources
in the field having different count rates in both detectors as a
result of different spectral responses.

Counts were extracted in the 2-7 keV band (PI channels 548-1708) since
above 7~keV the detector background count rate is larger than the
cosmic XRB count rate. The count rate to flux conversion factors,
however, were computed for the standard 2-10~keV band, assuming a
single power law spectrum, and energy spectral index of 0.7.

\begin{figure}
\vbox to 0cm{\vfil}
\epsfverbosetrue
\epsfysize=210pt
\epsffile{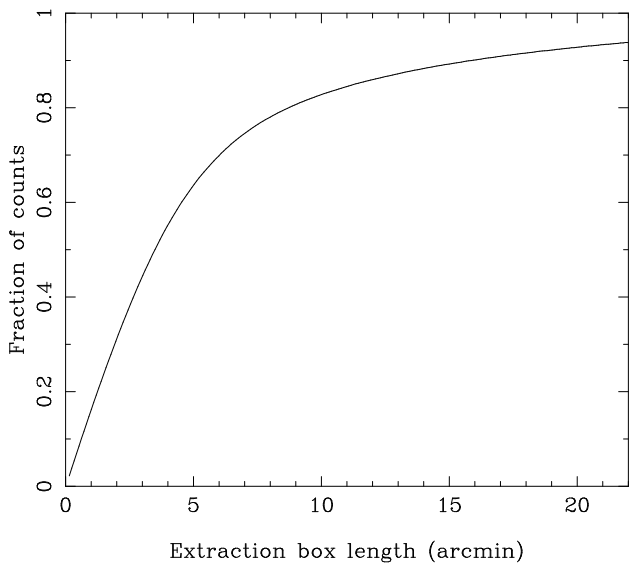}
\caption{Fraction of the counts that are collected in a square
extraction box from a point source at its center as a function of box 
side length.}
\end{figure}

The data were extracted in square spatial bins corresponding to a
scale $5.6 \times 5.6\, {\rm arcmin}^2$, so each one of the 4 CCDs in
SIS0 was divided into 4 extraction bins.  The reason for this choice
results from a trade-off between having the largest possible number
of data measurements without making the neighbouring pixels too
strongly dependent.  

To further emphasize this point, we have carried out simulations of
the PSF using a ray-tracing routine with improved optical constants
(see Section 3.2 for more details). In Fig.~2 we show the fraction of
counts from a point source collected in a centered square bin as a
function of side length.  It is evident that there is an inflection at
around $5\, {\rm arcmin}$ out to which about 60 per cent of the counts
have been collected and beyond which the counts are much more spread
over the whole image.  Indeed, taking a box of, say, $3\times 3\, {\rm
arcmin}$ would make the neighbouring extraction bins highly dependent
with only $\sim 40$ per cent of the energy collected inside the box.

Therefore, a total of 240 measurements of the XRB intensity in
different sky positions have been
extracted. The distribution of these intensities (Fig.~3) does not
only reflect confusion P(D) noise fluctuations, but also some
additional broadening due to the different
sensitivities in the 16 extraction regions in each image. 

\begin{figure}
\vbox to 0cm{\vfil}
\epsfverbosetrue
\epsfysize=210pt
\epsffile{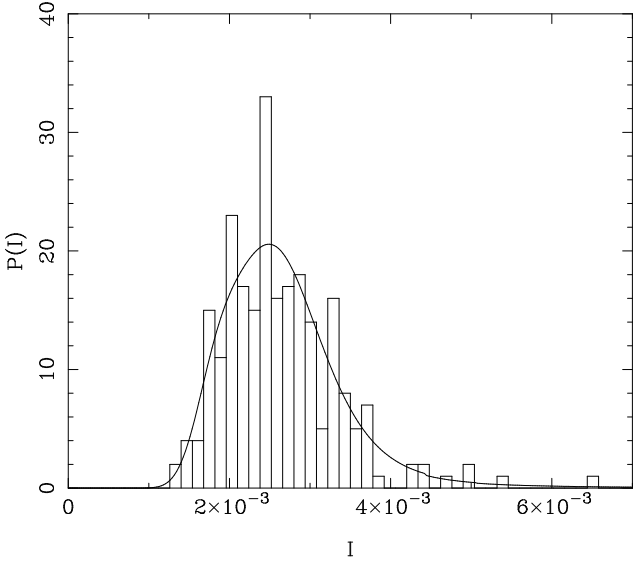}
\caption{The distribution of measured intensities together with the
best fit according to the models discussed in Sections 3 and 4.}
\end{figure}

\section {Modelling the fluctuations}

\subsection {Fluctuations and source counts}

The basic theory that relates the expected fluctuations in the XRB in
terms of the source counts (Scheuer 1974, Condon 1974) needs to be
specifically adapted to the study of the ASCA fluctuations.  This is a
particularly complicated situation because of two reasons.  The first
of them is that the PSF is very extended, with significant wings that
reach outside the detector (see Fig.~2). The second one is that the
light collected by the SIS0 CCDs comes not only from the sources
nominally within its field of view, but sources out to $\sim 2\deg$
off the optical axis can still produce a substantial contribution.  In
fact, some vignetting was expected within the SIS field of view which is
not observed, the reason being the influence of the outside sources.

We then consider a $5.6\times 5.6\, {\rm arcmin}^2$ bin centered at
point $\vec x_b$ (tangential coordinates). If we assume a homogeneous
distribution of $N$ sources in the sky (at positions $\vec x_i$),
whose fluxes $S_i$ are distributed according to the differential
source counts $N(S)$ (sources per unit flux per unit solid angle), the
intensity (in counts per second) collected at this particular bin is
\begin{equation}
I(\vec x_b)=\sum_{i=1}^{N}S_i\, F_1(\vec x_b,
\vec x_i) F_2(\vec x_b)
\end{equation}
where $F_1(\vec x_b,\vec x_i)$ is the rate of photons per second
that would land inside the extraction box centered at $\vec x_b$ when
a source of unit flux is placed at a point $\vec x_i$ in the sky, and
$F_2(\vec x_b)$ is the detector quantum efficiency in that box.

Here the `conversion factor', for the extraction box centered at $\vec
x_b$, is defined as the count rate detected in that box of the CCD
from a source of unit flux at the centre of the box ($\vec x_b$),
i.e.,
\begin{equation}
C(\vec x_b)=F_1(\vec x_b, \vec x_b)F_2(\vec x_b)
\end{equation}

The function $F_1(\vec x_b, \vec x)$ can also be regarded, up to a
multiplicative constant, as the
sensitivity function of this particular extraction box to a point in
the sky $\vec x$.  Following Condon (1974), the sensitivity function,
normalised in such a way that its maximum value is unity, is related to
$F_1$ as follows
\begin{equation}
f(\vec x_b,\vec x)={F_1(\vec x_b, \vec x)\over F_1(\vec x_b, \vec
x_b)}
\end{equation}
and therefore eq. (1) can be re-written as 
\begin{equation}
I(\vec x_b)=\sum_{i=1}^{N} S_i\, C(\vec x_b)\, f(\vec x_b, \vec x_i).
\end{equation}

The distribution of intensities, for this particular extraction
region, can be expressed in the usual Fourier transform terms:
\begin{equation}
P(I;\vec x_b)=\int
d\omega \exp (-2\pi i\omega I) \exp (\Psi (\vec x_b,\omega))
\end{equation}
where
\begin{equation}
\Psi(\vec x_b,\omega)=\int d^2 x\, \int dS N(S) \left[ {\rm e}^{2\pi i \omega
S\, C(\vec x_b)\, f(\vec x_b,\vec x)}-1\right]
\end{equation}
This distribution needs to be convolved with photon counting noise
before it can be compared to data.  Thus the probability of measuring
$N_c$ counts is given by
\begin{equation}
P_c(N_c,\vec x_b)=\int dI\, P(I,\vec x_b){\cal P}\left(N_c,\left(I+I_{det}\left(\vec x_b\right)\right)t\right)
\end{equation}
where ${\cal P}(N_c,n_c)$ is the Poisson probability of measuring
$N_c$ counts, the mean being $n_c$. $I_{det}(\vec x_b)$ is the
detector (non-X-ray) background count rate, which can be measured via
dark Earth observations and $t$ is the exposure time.

The basic ingredients to compute eqs.(5) and (7) are then the source
counts, the sensitivity function of that extraction box and the
conversion factor defined in eq.~(2) for that extraction box. Note
that these last two depend on the particular extraction box and that the
conversion factor in addition depends on the time when the data
were taken (see Fig.~1 and discussion in Section 2).

\begin{figure}
\vbox to 0cm{\vfil}
\epsfverbosetrue
\epsfysize=210pt
\epsffile{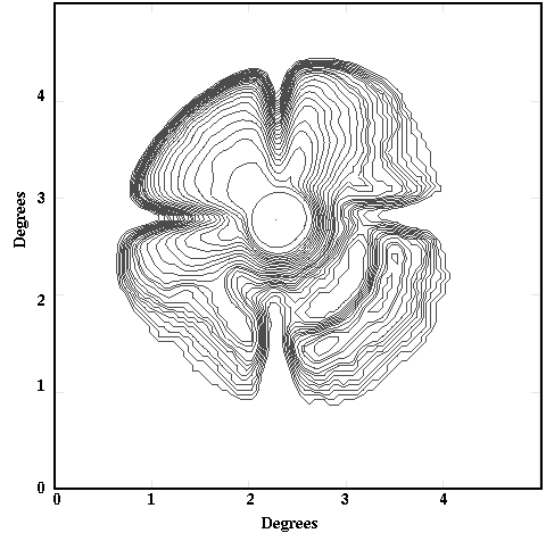}
\caption{Sensitivity map of one of the extraction regions. 
The 30 contours are logarithmically spaced in sensitivity from $3.9\times
10^{-1}$ to $2.2\times 10^{-4}$.}
\end{figure}

\subsection {Sensitivity maps and conversion factors}

In order to build up sensitivity maps (i.e. the functions $f(\vec x_b,
\vec x)$ for each of the 16 values of $\vec x_b$) we have carried out
simulations, using a ray-tracing routine, to study the properties of
the X-ray telescope (XRT) that focuses the X-rays.  The details of the
ray-tracing code can be found in Kunieda, Furuzawa \& Watanabe (1995)
and Gendreau \& Yaqoob (1997).  Basically it takes every photon, whose
energy and incidence angle are known, and does a Monte Carlo
simulation of its possible trajectories, including reflection in the
telescope mirrors, absorption, multiple scatterings etc... The
parameters governing the various processes have been updated after
experience has been accumulated on the behaviour of the XRT.

In order to construct the sensitivity map for each extraction region,
a uniform distribution of sources has been simulated in a
$4^{\circ}\times 4^{\circ}$ region around the optical axis. Each
source was assigned a flux according to a $N(S)$ (although this is
irrelevant for this specific purpose). The number of photons coming
from each source was computed assuming a geometric area of $882 {\rm
cm}^2$ for the XRT and a source spectrum with an energy spectral index
of $0.7$. That produced a list of photons with their corresponding
energies and incident positions in the sky. These photon lists were
then ray-traced using the above code. 

Since each photon carried labels with its original incoming direction,
it was possible to build a sensitivity map.  The incoming positions in
the sky of the photons collected in
each one of the particular extraction boxes were recorded.  The density of
these positions in the sky, properly normalized, is the sensitivity function.

Fig.~4 shows a
contour diagram with the sensitivity map of one of the 16 extraction
regions. In that map it can be seen that there is sensitivity out to
very large offset angles.  In the same figure it can also be noted
that the sensitivity map has strong gradients, particularly close to
the ``cross''. (This shape results from the quadrant construction of
the telescope).

\begin{figure}
\vbox to 0cm{\vfil}
\epsfverbosetrue
\epsfysize=210pt
\epsffile{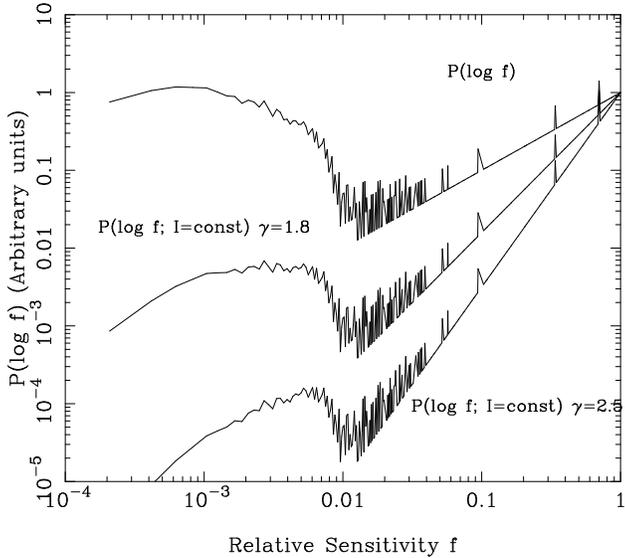}
\caption{The top curve shows the distribution in $\log f$, $f$ being
the sensitivity in one of the extraction regions, obtained from the
ray-tracing simulations. The two bottom curves show the same
distribution but for a fixed constant value of the intensity produced
by a single source, assuming a single power-law model for the source
counts (see text for details). A horizontal line would represent equal
contribution from each decade in sensitivity.}
\end{figure}

To emphasize the effect of the extent of the sensitivity functions,
Fig.~5 shows the distribution function for the sensitivities in a
specific spatial bin $f$ when smoothed on 1~arcmin$^2$ bins. The
contribution to the total intensity received in a given spatial bin by
sources in different directions is weighted by this distribution (top
line in Fig.~5).  Indeed, for a uniform distribution of sources
$P(\log f)$ measures the relative contribution to this total intensity
from different values of $\log f$.  Fig.~5 emphasizes that the
contribution from markedly off-axis sources ($f\sim 10^{-3}$) is at
least as important as the one from the on-axis sources ($f\sim 1$)
when computing the total intensity.  Therefore stray light has to be
accurately modelled if an absolute measurement of the XRB intensity
needs to be done.

To study whether the XRB fluctuations are also dominated by off-axis
sources, we recall that these fluctuations will be dominated by a very
limited range of intensities, corresponding to the dispersion in the
distribution of the fluctuations. We then construct the bivariate
distribution of intensity $I$ and sensitivity $\log f$ for a given
spatial bin.  The intensity produced by a single source with flux $S$
in a particular bin is $I=f\, S$. Using a single power-law form for
the differential source counts with slope $\gamma$ the bivariate
function for $I$ and $\log f$ is
\begin{equation}
P(\log f;I)\propto I^{-\gamma} f^{\gamma-1} P(\log f)
\end{equation}
Fig.~5 also shows this distribution for constant $I$ and different
values for $\gamma$ (lower curves) for an arbitrary constant value of
the intensity $I$.  Therefore, the contribution to the XRB
fluctuations (which are dominated by a specific value of $I$) from
sources at different positions in the sky (i.e., different
sensitivities) is measured by this bivariate distribution for constant
$I$. Fig.~5 emphasizes that for any reasonable slope of the source
counts, the fluctuations will always be dominated by sources within
the nominal field-of-view of the corresponding spatial bin ($f\sim
1$). Therefore, stray light does not dominate the XRB fluctuations in
this case, although it is very important for the computation of the
average intensity received in each spatial bin.

In order to perform the integration of eq~(6), the sensitivity
functions have been computed with 1~arcmin$^2$ resolution (as in
Fig.~5) which represents a compromise between good statistics in the
extended low sensitivity tails and smoothing the regions with strong
gradients.

With the ray-tracing simulations we also computed the conversion
factors introduced in eq.~(2). A bright source was placed in the
middle of each bin, its photons were ray traced and only those which
landed on the bin itself were counted.  A detector efficiency model
was used (Gendreau 1995) to convert from photons to counts (i.e.,
the function $F_2(\vec x_b)$.  The RDD effects discussed in section 2
were incorporated into that detector efficiency model both to estimate
the conversion factor $C(\vec x_b)$ and the non-X-ray detector
background $I_{det}(\vec x_b)$.

\section{Results}

\subsection{The fitting process}

Given that each one of the 16 extraction regions has a different
sensitivity function and a different conversion factor, and that RDD
changes the conversion factor for each observation, a model for the
XRB fluctuations had to be computed for each one of the 240 data
points, for every set of values of parameter space.  For every set of
values in parameter space a model
distribution (eqs. 5-7) was computed for each one of the measured
intensities, taking into account the different sensitivity functions
(this is different for each one of the 16 extraction regions),
the different conversion factors (for a given extraction region it
also changes from image to image due to the RDD) and different
exposure times.    
Particular
attention was paid to the fact that the mean intensity expected for
each data point is not known with infinite accuracy, due to the
statistical uncertainty in the measurement of the cosmic XRB, and also
possible overall inaccuracies in the conversion factors or sensitivity
functions.  This is why we prefer to fit the mean value of the
intensity rather than impose it.  The mean expected value for a
given data point is 
\begin{equation}
\langle I(\vec x_b)\rangle = C(\vec x_b) \Omega (\vec x_b) \int_{S_{min}}^{S_{max}} dS\,
N(S)+I_{det}(\vec x_b)
\end{equation}
where
\begin{equation}
\Omega(\vec x_b)=\int d^2x\, f(\vec x_b,\vec x)
\end{equation}
is the effective solid angle for flux collection.  Here we take
$S_{min}=3\times 10^{-11}\, \erg$ (which is the completeness limit of
the Piccinotti et al sample, and whose sources were certainly not near
the ASCA images) and $S_{min}$ is either the flux at which the XRB
saturates or $10^{-15}\, \erg$ (small enough so that fluctuations
produced by fainter fluxes will be absolutely negligible) when the
slope of the source counts is too flat to saturate the XRB. Then, the
quantity
\begin{equation}
X={I(\vec x_b)-I_{det}(\vec x_b)\over C(\vec x_b)\Omega(\vec x_b)}
\end{equation}
should not depend on extraction box or detector efficiency and reflect
only sky fluctuations and counting noise.  We evaluated the average of
this quantity $\langle X\rangle_{obs}$ and its uncertainty
$\sigma(\langle X\rangle)_{obs}$ using the 240 data points
themselves. In the fitting process, however, $\langle X\rangle$ was a
free parameter and was fitted (and considered non-interesting) but
adding the following contribution to the $\chi^2$ to reflect our
knowledge on its value, i.e.
\begin{equation}
\left({\langle X\rangle-\langle X\rangle_{obs}\over \sigma(\langle
X\rangle)_{obs}}\right)^2.
\end{equation}

A number of tests were performed to check on the accuracy of our
fitting method (and model) by generating simulated images with model
source counts and then analyzing them with the same procedure that was
later applied to the data.  Both a maximum likelihood method and a
$\chi^2$ method (where the intensity histogram was binned in groups of
at least 15 data points to ensure proper statistics) were tried.  As a
general result, the maximum likelihood method almost invariably found
much higher source counts than the input ones, especially when bright
sources were present. The $\chi^2$ minimisation proved rather accurate
in defining the value of the source counts in the range of fluxes
where this analysis is sensitive ($\sim (2-13)\times 10^{-14}\erg$ for
the real data), although the shape of the source counts within this
flux range is very poorly constrained by a sample of this size and
depth.

\subsection{Single power law source count models}

We assume that the source counts follow a power-law distribution
\begin{equation}
N(S)={K\over S_0}(\gamma-1)\left({S\over S_0}\right)^{-\gamma}
\end{equation} where the reference flux $S_0$ is chosen at
$10^{-13}\erg$ (which is close to the flux where the fluctuation
analysis is most sensitive) and $K$ is the number of sources per solid angle
brighter than $S_0$. Results from GINGA fluctuation analyses (Butcher
et al 1997) predict values of $K=400\, \deg^{-2}$ and $\gamma\approx
2.5$, if they can be extrapolated to these lower fluxes. On the
contrary, if a conversion from the ROSAT source counts in the
0.5-2~keV band to our 2-10~keV band in terms of a single power law
X-ray spectrum with energy spectral index of $0.7$ is assumed, then a
normalisation closer to $K=150\, \deg^{-2}$ would apply.

\begin{figure}
\vbox to 0cm{\vfil}
\epsfverbosetrue
\epsfysize=210pt
\epsffile{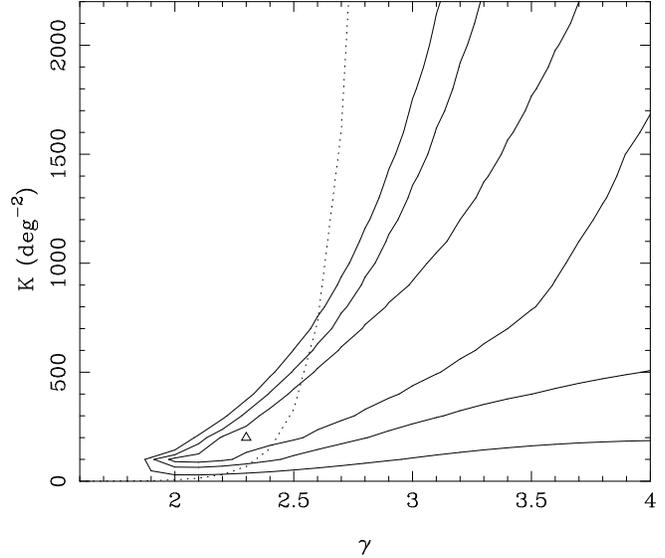}
\caption{Best fit and confidence regions for 1, 2 and 3 $\sigma$ for a
single power law model. The dotted line shows those points which will
be consistent with the HEAO-1 A2 point of Piccinotti et al (1982).}
\end{figure}

\begin{figure*}
\vbox to 0cm{\vfil}
\epsfverbosetrue
\epsfysize=280pt
\epsffile{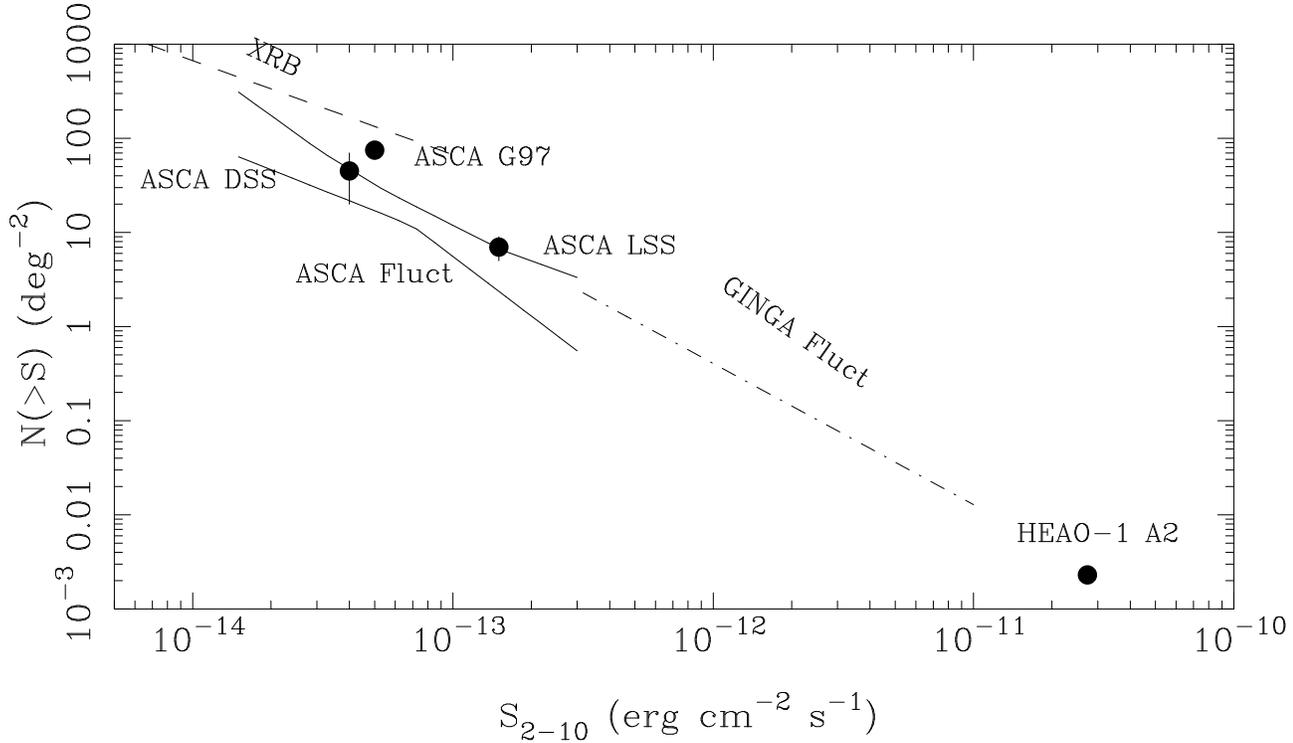}
\caption{The 2-10 keV integral source counts. The data points are from:
A2 by Piccinotti et al (1982), LSS by Ueda (1995), DSS by Inoue et al (1996),
G97 by Georgantopoulos et al (1997). The `trumpet' is the 1$\sigma$
region delimited by our analysis. The dashed line shows the saturation
level for the XRB assuming $\gamma=2.5$.}
\end{figure*}

The results from the fitting process are shown in Fig.~6 in
($K$,$\gamma$) space. The best fit corresponds to $K\approx
200\, \deg^{-2}$ and $\gamma\approx 2.3$ with a $\chi^2=17.33$ for 13
degrees of freedom (16 bins minus 3 fitted parameters) resulting in a
fairly small reduced $\chi^2/N_{\rm dof}\sim 1.3$. There is, however, no
obvious systematic difference between the histogrammed data and the
best-fit model (see again Fig.~3).

In Fig.~6 it can also be seen that there is a degeneracy
between values of $K$ and $\gamma$ which are very poorly constrained
individually.  When $K$ is considered the only interesting parameter,
it can take any value from 100 to 750 $\deg^{-2}$ (1$\sigma$
interval). A similar analysis for $\gamma$ yields a 1$\sigma$ interval
that ranges from $\gamma\sim 2$ to $\gamma\sim 3$.

However, if the source counts are forced to have an euclidean shape,
then the normalisation is fairly well constrained to $K=(300\pm 100)\,
\deg^{-2}$ (1$\sigma$ errors), the ROSAT normalisation $K=150\,
\deg^{-2}$ being only 2$\sigma$ below in this case.

In spite of the fact that the parameters $K$ and $\gamma$ are poorly
constrained individually, as expected from the simulations the source
counts in the region between $(2-12)\times 10^{-14}\erg$ are fairly
well constrained by our analysis. Fig.~7 shows this result in terms of
the integral source counts together with the results of the various
surveys (errors are 1$\sigma$ always). Our result is consistent with
the ASCA LSS (Ueda 1995) and the ASCA DSS (Inoue et al 1996), but the
Georgantopoulos et al (1997) source counts down to $5\times
10^{-14}\erg$ are significantly higher than what we find in our
analysis.  In fact they are about 3$\sigma$ above from our result.
The reasons for this discrepancy (and also the disagreement between
Georgantopoulos et al 1997 and the ASCA DSS) are unclear, but small
number statistics, source variability, confusion and Eddington bias
are among the possibilities. In particular, as recognised by
Georgantopoulos et al (1997), Eddington bias and confusion could well
produce errors of the order of 100 per cent in the fluxes of the
faintest sources.

A slight modification applies to our source counts if the source
spectra are flatter.  If an energy spectral index of $0.4$ (similar to
that of the XRB) is assumed instead of the canonical $0.7$, larger
fluxes are needed to produce the same intensities due to the dominant
response of the XRT+SIS0 at low energies.  In this case, our estimates
of the source counts shown in Fig.~7 would have to be displaced to
higher fluxes by rougly 15 per cent which is the average ratio between
conversion factors (as defined in eq.~2) for energy spectral indices
0.4 to 0.7.  For the approximately euclidean source counts, this means
that our source counts will go up by $\sim 25$ per cent, still
significantly lower than the Georgantopoulos et al (1997) source
counts.

\subsection{Broken power law source counts}

As discussed earlier, soft X-ray source counts found by ROSAT exhibit
a break from an approximately euclidean slope above a 0.5-2~keV flux
of $2\times 10^{-14}\erg$. Converting that break flux into the 2-10
keV band requires detailed knowledge of the broad-band average X-ray
spectrum of the sources at that flux, which is not known.
Alternatively, if a break in the 2-10~keV source counts is found and
is identified with the ROSAT one, an approximate conversion factor
between the 0.5-2 and 2-10 keV bands could be found.

Therefore, we tested also a broken-power law source counts model, which fits
the 2-10~keV source counts at bright fluxes (i.e., eq.(13) with
$K=400\, \deg^{-2}$ and $\gamma=2.5$) down to a break flux $S_b$ below
which the source counts flatten to a slope $\gamma$. After modelling
the source counts in terms of these two free parameters, it was found
that all values of $S_b$ were within 1$\sigma$ of the best fit for
$\gamma < 2.5$ and $S_b>3\times 10^{-14}\erg$.

\begin{figure}
\vbox to 0cm{\vfil}
\epsfverbosetrue
\epsfysize=210pt
\epsffile{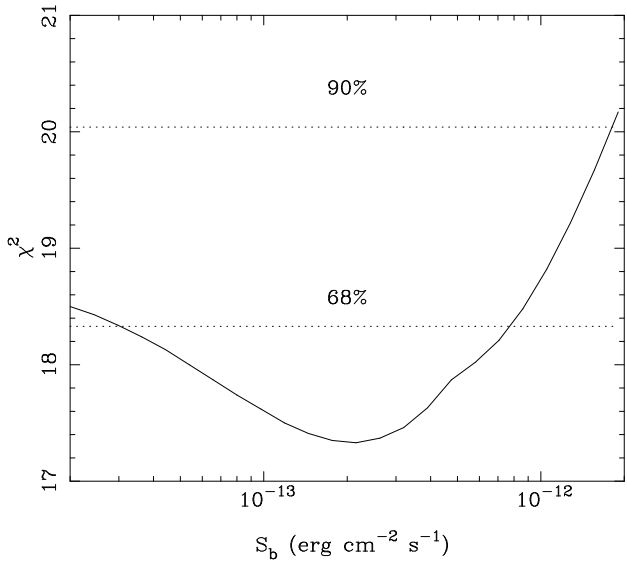}
\caption{$\chi^2$ for the break flux $S_b$ in a broken power law model
where the slope below $S_b$ has been taken $\gamma=1.8$.}
\end{figure}

A further test was carried out, by fixing the slope below the break to
the value found by ROSAT ($\gamma=1.8$). $\chi^2$ as a function of
the break flux $S_b$ is plotted in Fig.~8. The best fit is $\log
(S_b)=-12.7^{+0.5}_{-0.8}$ (1$\sigma$). The 90 per cent confidence
upper limit is $\log (S_b)< -11.7$.  Unfortunately these limits do not
impose any relevant constraints on the broad band spectrum of the
sources, which is only restricted to have an energy spectral index
of less than 0.9 ($1\sigma$).

\section{Discussion and future work}

In this paper we have found that the extrapolation of euclidean
power-law source counts from higher fluxes appears to be consistent
with the fluctuations in deep ASCA SIS0 images down to a flux of $\sim
2\times 10^{-14}\erg$.   Adopting the euclidean form, there must
be $(300\pm 100)\, \deg^{-2}$ sources brighter than a 2-10~keV flux of
$10^{-14}\erg$. Down to the flux where our fluctuation analysis is
sensitive ($2\times 10^{-14}\erg$) the integrated intensity of the
sources represents $35\pm 13$ per cent of the XRB as measured by the
ASCA SIS (Gendreau et al 1995). There is no evidence for a break in
the source counts although it cannot be excluded. For reasonable
source spectra, the break that is seen in soft X-ray source counts is
expected to occur around the faintest flux at which the fluctuation
analysis is sensitive, so it is difficult to detect.

Although the analysis of the fluctuations in a telescope which has
such an extended sensitivity is complicated, we have shown that the
analysis can be done if all effects are properly modelled.  The major
limitation in our results comes from the limited field of view of the
SIS0 detector.  The next step would be to use GIS images of the same
fields to increase by a factor of several the solid angle surveyed
and, hopefully, reduce the uncertainties in the determination of the
source counts by a factor of at least 2.

Another issue that can be addressed with the fluctuation analyses is
that of the average spectrum of the sources.  Given the fairly good
spectral resolution of both the SIS and the GIS, fluctuations in two
different energy bands can be fitted to the same source counts, the
major unknown being the spectral shape linking both bands.  Modelling
the spectrum in terms of a single power law, will allow the average energy
spectral index of the sources in the flux range $(2-12)\times
10^{-14}\erg$ being determined.  That particular approach proved
useful in deriving the soft X-ray spectrum of very faint sources by
analysing the fluctuations in ROSAT PSPC deep images (Ceballos, Barcons
\& Carrera 1997).

\section*{acknowledgments}

We thank Richard Mushotzky and Keith Jahoda for helpful comments.
This research has made use of the HEASARC archive which is maintained
by NASA at GSFC.  XB acknowledges partial financial support provided
by the DGES under project PB95-0122 and funding for his sabbatical at
Cambridge under DGES grant PR95-490. ACF thanks the Royal Society for
support.

\bsp 

\label{lastpage}


\begin{thebibliography}{99}


\bibitem{b01} Almaini, O., Shanks, T., Boyle, B.J., Griffiths, R.E.,
Roche, N., Stewart, G.C., Georgantopoulos, I., 1996, MNRAS, 282, 295
\bibitem{b02} Barcons, X.,  Fabian, A.C., 1990, MNRAS, 243, 366
\bibitem{b03} Barcons, X., Branduardi-Raymont, G., Warwick, R.S.,
Mason, K.O., McHardy, I.M., Rowan-Robinson, M., 1994, MNRAS, 268, 833 
\bibitem{b04} Barcons, X.,  Fabian, A.C., Carrera, F.J., 1997, MNRAS,
in the press (astro-ph/9705180)
\bibitem{b05} Branduardi-Raymont, G. et al 1994, MNRAS, 270, 947
\bibitem{b06} Butcher, J.A., Stewart, G.C., Warwick, R.S., Fabian,
A.C., Carrera, F.J., Barcons, X., Hayashida, K., Inoue, H., Kii, T.,
1997, MNRAS, in the press (astro-ph/9707135)
\bibitem{b07} Ceballos, M.T.,  Barcons, X., Carrera, F.J., 1997,
MNRAS, 286, 158
\bibitem{b08} Condon, J.J., 1974, ApJ, 188, 279
\bibitem{b09} Dotani, T., Yamashita, A., Rasmussen, A., 1995, ASCA
News, 3, 25
\bibitem{b10} Fabian, A.C.,  Barcons, X.,  1992, ARA\&A, 30, 429
\bibitem{b11} Gendreau, K.C., 1995, PASJ, 47, L5
\bibitem{b12} Gendreau, K.C., Yaqoob, T., 1997, ASCA News 5, 8
\bibitem{b13} Gendreau, K.C., 1995, Ph.D. Thesis, MIT
\bibitem{b14} Georgantopoulos, I., Stewart, G.C., Blair, A.J., Shanks,
T., Griffiths, R.E., Boyle, B.J., Almaini, O., Roche, N., MNRAS, in
the press (astro-ph/9704147)
\bibitem{b15} Hamilton, T.T.,  Helfand, D.J., 1987, ApJ, 318, 93
\bibitem{b16} Hasinger, G., Burg, R., Giacconi, R., Hartner, G.,
Schmidt, M., Tr\"umper, J., Zamorani, G., 1993, A\&A, 175, 1
\bibitem{b17} Hasinger G., Burg R., Giacconi R., Schmidt M., Tr\"umper
J., Zamorani G., 1997, A\&A, in the press (astro-ph/9709142)
\bibitem{b18} Hayashida, K., 1989, Ph.D. Thesis, University of Kyoto
\bibitem{b19} Inoue, H., Kii, T., Ogasaka, Y., Takahashi, T., Ueda,
Y., 1996, MPE Report, 263, 323
\bibitem{b20} Kondo, H., 1991, Ph.D. Thesis, University of Tokyo
\bibitem{b21} Kunieda, H., Furuzawa, A., Watanabe, M., 1995, ASCA
News, 3, 3
\bibitem{b22} McHardy, I.M., et al 1997, MNRAS, in the press (astro-ph/9703163)
\bibitem{b23} Piccinotti, G., Mushotzky, R.F., Boldt, E.A., Holt,
S.S., Marshall, F.E., Serlemitsos, P.J., Shafer, R.A., 1982, ApJ, 253,
485
\bibitem{b24} Romero-Colmenero, E., Branduardi-Raymont, G., Carrera,
F.J., Jones, L.R., Mason, K.O., McHardy, I.M., Mittaz, J.P.D., 1996,
MNRAS, 282, 94
\bibitem{b25} Scheuer, P.A.G., 1974, MNRAS, 166, 329
\bibitem{b26} Schmidt M. et al, 1997, A\&A, in the press (astro-ph/9709144)
\bibitem{b27} Tanaka, Y., Inoue, H., Holt, S.S., 1994, PASJ, 46, L37
\bibitem{b28} Ueda, Y., 1995, Ph. D. Thesis, University of Tokyo

\end{thebibliography}
\end{document}